\documentclass[12pt]{article}
\usepackage{graphicx}		
\begin{document}
\title{An integrable evolution equation for surface waves in deep water}
\author{R. Kraenkel$^a$, H. Leblond$^b$ and  M. A. Manna$^c$\\ 
$^a$ Instituto de F\'isica T\'eorica (UNESP)\\
Universidade Estadual Paulista (UNESP),\\
Rua Dr. Bento Teobaldo Ferraz 271\\
Bloco II, 01140-070, S\~ao Paulo, Brazil\\
$^b$Laboratoire de Photonique d'Angers,\\
Universit\'e d'Angers, 2 Bd Lavoisier 49045 Angers Cedex 1,\\
and\\
$^c$ Laboratoire Charles Coulomb, \\UMR 5221 CNRS-UM2
Universit\'e  Montpellier II, \\F-34095 Montpellier Cedex 5 - France
\footnote{Principal corresponding author}}
\maketitle
\begin{abstract}
In order to describe the dynamics of monochromatic surface waves in deep water,
we derive a
nonlinear and dispersive system of equations for the free surface elevation and 
the free
surface velocity from the Euler equations in infinite depth. From it,
and using a multiscale perturbative methods, an asymptotic model for 
small-aspect-ratio waves is derived. The model is shown to be completely 
integrable. The Lax 
pair, the first conserved quantities as well as the symmetries are exhibited. 
Theoretical and  numerical studies  reveal that it supports periodic 
progressive Stokes waves which peak and break in finite time. Comparison
between the limiting wave solution of the asymptotic model and classical 
irrotational results is performed. 
\end{abstract}
%--------------------------------------------
\section{Introduction}
\label{sec:introduction}
%--------------------------------------------
 Wave propagation  in an ideal incompressible fluid is a classical
matter of investigation in mathematical physics. Especially, surface
gravity waves have been intensively studied and many model equations
were introduced to handle this problem. 
Such studies are in a large part motivated by the fact that the initial
three-dimensional water wave problem is  not tractable analytically.
Traditionally, the subject was studied in two almost separated domains: 
{\it shallow water models} and {\it deep water models}.

The shallow water theory is widely known. The irrotational or rotational Euler equations 
have been approximated from several sides. An approach is based on the use of two parameters $\alpha$ and $\beta$. 
$\alpha = a/h $  measures the amplitude $a$ of the perturbation with respect 
to the depth $h$ and  $\beta = h^2/\lambda^2 $ measures the depth with respect  to 
the wavelength $\lambda$. Assuming $\alpha$ and $\beta$ smaller than one, a
perturbative procedure is carried out. Then model equations are obtained
by retaining only the lowest order terms in $\alpha $ and $\beta$  \cite{whitham}. 
Another well-established approach for the  derivation of 
shallow water model equations is the reductive perturbation method \cite{taniuti1, taniuti2, taniuti3,tutorial} 
based on the Gardner-Morikawa transformation \cite{gardner}. It enables us to introduce slow space and time variables 
able to
describe the effect of nonlinearity and dispersion asymptotically in space and
time. An alternative method was introduced by Serre \cite{Serre} and several
years later by Su and  Gardner \cite{gardner} and Green and Naghdi
\cite{GLN1, GLN2}. It is based on an Ansatz which uses the shallow water limit of the
exact linear solution of the rotational Euler equations. This way were obtained the 
Nonlinear Shallow Water \cite{whitham}, the Boussinesq   
\cite{boussinesq, boussinesqmodels}, the  Korteweg-de Vries (KdV), the modified KdV 
(mKdV) \cite{korteweg,kortewegmodel}, the Kadomtsev-Petviashvili 
\cite{kape}, the Benjamin-Bona-Mahony-Peregrine \cite{bbmp}, the Serre, the  Green-Naghdi and more recently the
Camassa-Holm \cite{CH}  equations. All these model equations govern the
asymptotic dynamics of \textit{wave-profiles of long-waves in shallow water}.

The deep water  limit consists in considering an infinite depth ($h=-\infty$) in the Euler equations. Then neither $\alpha$ nor $\beta$ % nor the Gardner-Morikawa transformation
can be defined. The Gardner-Morikawa transformation, based on the 
dispersionless shallow water linear limit of the Euler equations, cannot be 
defined any more because the deep water limit is fully dispersive.
Consequently, surface wave propagation in deep water is mainly concerned with the 
{\it nonlinear modulation of wave trains}. The most
representative model equation is the ubiquitous NonLinear Schr\"odinger equation (NLS)
\cite{kortewegmodel,zakharov,ablowitz}. For a full account on modulation of short wave trains
in water of intermediate or great depth see Mei \cite{mei}.
However, some model equations have been given for wave profile evolution in deep water. E.g.,
 in reference \cite{matsuno} is derived a finite-depth Boussinesq-type 
equation for the profile of irrotational free surface waves. The derivation
is carried out via the theory of analytic functions followed by a perturbation procedure.
Its deep water limit is exhibited as a function of integral operators.
The Benjamin-Ono equation, first introduced by T. Benjamin \cite{Benjamin} and H. Ono 
\cite{Ono}, is worth being mentioned. It is a nonlinear partial 
integro-differential equation which describes one-dimensional internal wave
profile evolution in deep water. In reference \cite{leon} an  approach analogous
the one we develop in the present paper was introduced.
  Other model equations whose solutions behave as deep water waves can be found in
references \cite{whitham2} and \cite{bro}. Their dispersion relations coincide
exactly with that of  water waves in infinitely deep water and
the nonlinear terms are chosen in a \textit{ad hoc} way in order to reproduce the Stokes limiting
wave.

The present work is the extension of a long sequel of studies carried out by the
authors in various physical contexts, dealing with model equations for nonlinear and dispersive 
short wave dynamics (\cite{short1}-\cite{short15}).
 Especially,
the purpose of the paper is to study the dynamics of an
elementary wave profile in deep water. Instead of looking for  modulation
dynamics of a wave train, we seek for the combined effects of
dispersion and nonlinearity on a given \textit{ Fourier component} with wave vector $k$. 
Our approach is a generalization of the method firstly used by Serre for
surface waves in shallow water. We assume an Ansatz  
based on the exact linear solution of the Euler system for deep water.
Of course, as in the shallow water case, the Ansatz does not yield an exact solution
of the full Euler system. It produces a \textit{depth-average model equation}, which is expected to
be valid close to the linear limit of the Euler system. 

The paper is organized as follows. In Section (\ref{sec:euler}) we introduce 
the Euler equations in deep water. In Section (\ref{sec:linear pattern}), after 
a brief summary of the "modus operandi" of the linear pattern Ansatz in 
the shallow water limit, we 
generalize it to the deep water case. In Section 
(\ref{sec: water wave problem}) we give the nondimensionalisation of the Euler 
equations and we introduce the deep water surface wave problem. In Section
(\ref{sec:serre}) we
introduce the Serre approximation in the deep water context and we derive
a nonlinear and dispersive system of equations for the free surface elevation
and for the free surface velocity. In Section (\ref{sec:small aspect}) we derive a 
small-aspect-ratio wave equation in deep water. 
In Section (\ref{sec:Mathematical}) are exhibited the mathematical
properties of the model: the Lax pair, the relation with the integrable 
Bullough-Dodd model and the symmetries. In  Section (\ref{sec:Stokes}) is studied the 
progressive periodic wave solution and the limiting wave. 
The work is completed with a numerical study of the
limiting wave. In the Appendix we give some detail on the derivation of the
Bullough-Dodd equation and finally Section  (\ref{sec:conclusion}) draws 
the conclusions.
%--------------------------------------------
\section{The Euler equations in deep water}
\label{sec:euler}
%--------------------------------------------
We considerer the   dimensional Euler equations for deep water.
Let the particles of the fluid be located relative to a fix rectangular 
Cartesian frame with origin $O$ and axes $(x, y, z)$, where 
$Oz$ is the 
upward vertical direction. We assume translational symmetry along $y$ 
and we will only consider a sheet of fluid parallel to the $xz$ plane.
The velocity of the fluid is $\vec V(x,z,t) = (U(x,z,t), W(x,z,t))$. 
The fluid sheet is moving on a  bottom at $z = -\infty$, and its 
upper free surface is located at $z =\eta (x,t) $.  
The continuity equation and the Newton 
equations (in the flow domain)  read as
\begin{eqnarray} 
U_x + W_z &=& 0 \; , \label{cont}\\
\rho \dot{U} + P^{*}_{x}&=&0 \; ,  \label{newx}\\
\rho \dot{W} + P^{*}_{z} + g\rho &=&0,\; \label{newz}
\end{eqnarray}
in which subscripts denote partial derivatives, and the dot ( $\dot{ }$ ) denotes the material derivative defined by
\begin{equation}\label{matder}
\dot{F} = \frac{dF}{dt} = \frac{\partial F}{\partial t} + U
 \frac{\partial F}{\partial x} + W\frac{\partial F}{\partial z}\;.
\end{equation}
The boundary conditions at $z = -\infty$ and at $z = \eta (x,t)$ are
\begin{eqnarray}
W&=& 0 \;\; \mbox{for}\;\; z \rightarrow -\infty, \label{BC1}\\
P^{*}-P_0&=&0 \;\; \mbox{for} \;\;  z = \eta, \label{BC2}\\
\eta_{t} - W + U\eta_{x}&=&0   \;\; \mbox{for} \;\; 
 z = \eta\label{BC3}\;,
\end{eqnarray}
where $\rho$ is the constant density of the water, $g$ the  gravitation  constant and $P_0$
 the atmospheric pressure. The solution of the problem consists in finding
$\eta(x,t)$, $\vec{V}(x,z,t)$ and $P(x,z,t)$. 
\section{A deep water linear pattern Ansatz}
\label{sec:linear pattern}
Approximate solutions to the nonlinear surface water waves problem 
can be  computed only if the $z$ dependence of the  velocity field
$\vec{V}(x,z,t)$ is known. The $z$ dependence provides a coupling between the wave motion
at the surface and  the wave motion at various depths. In the linear case 
the $z$ dependence is known, hence the solution is exactly known. Expansions in power series of of $z$
as well as the theory of harmonic functions have been used to solve this
problem in the nonlinear and potential Euler equations. 
Another approach to the problem is the use of an Ansatz 
based on the linear
$z$ dependence of $\vec{V}(x,z,t)$. It is {\it the linear pattern Ansatz} 
introduced in references \cite{Serre, gardner, GLN1,GLN2} ,  
in the shallow water context, 
and equivalent to the widely known {\it columnar hypothesis}. It is thus worthy 
 to begin with a brief description of how the Ansatz works in the shallow 
water case, before we extend it to the deep water context.
 
{\it The shallow water Ansatz.}
For surface progressive waves of wave vector $k$ in water of depth $h$ in the shallow 
limit $kz\sim 0$, the linear solution $U_l(x,z,t)$ to the horizontal 
component $U(x,z,t)$ of the velocity and the dispersion relation $\omega$ are
\begin{eqnarray}
U_l(x,z,t)&=&A\exp{i(kx-\omega t)}\left[1 + 0(kz)^2\right],\label{shalU}\\
\omega(k)&=&k\sqrt{gh}\left[1 - \frac16(kh)^2 + ...\right],\label{shalomega}
\end{eqnarray}
where $A$ is a constant, in complex representation so that the physical velocity is given by real part of Eq. 
(\ref{shalU}). $U_l$ is a wave crest moving at the phase velocity
$c(k)$
given by
\begin{equation}
c(k)=\sqrt{gh}\left[1 - \frac16(kh)^2 + ...\right].
\end{equation}
The linear pattern Ansatz assumes that the $z$ dependence of $U(x,z,t)$
is the same as the $z$ dependence of $U_l(x,z,t)$, and that its spatiotemporal behavior is 
given by an undetermined function $u(x,t)$. So, Serre has assumed that 
\begin{equation}
U(x,z,t)=u(x,t),
\end{equation}
which corresponds to the extreme long wave limit in (\ref{shalU}), i.e, 
$U(x,z,t)$ does not depend on $z$.
 Since the phase velocity is invariant under 
$k \rightarrow -k$, i.e.,
\begin{equation}
c(k)=c(-k),
\end{equation}
the Ansatz does not need to take into account the direction of propagation in 
$u(x,t)$.

{\it The deep water Ansatz.}
We propose in the present paper an analogous procedure in the deep water case. Nevertheless two
important differences arise. The first one is that in the  
deep water 
case, according to Eqs.
(\ref{cont}-\ref{BC3}), we have
\begin{eqnarray}
U_l(x,z,t)&=&A\exp{i(kx-\Omega t)}\exp{kz},\label{deppUl}\\
\Omega(k)&=&\sqrt{gk}\label{deppOmega},\\
c(k)&=&\sqrt{\frac{g}{k}}\,.\label{deppc}
\end{eqnarray}
So, the phase velocity $c(k)$ is no longer invariant under $k\rightarrow -k$
and expressions (\ref{deppUl}), (\ref{deppOmega}) and  (\ref{deppc}) are strictly valid for wave crests
moving to the right only . Consequently, the Ansatz
would take into account the choice of the  propagation direction made
in deriving the linear solution.
The second difference is a very subtle one. Here we are interested in 
the effect of nonlinearity and dispersion on a purely sinusoidal wave of wave 
vector $k$ and frequency $\Omega$ for large $x$ and $t$.
However, wave vectors and frequencies are no longer constant in dispersive system, and are
 generalized to the concepts of local wave vector $k(x,t)$ and local
frequency $\Omega(x,t)$.
There will then be a non-uniform local wave train (nearly sinusoidal) 
with $k=k(x,t)$ and  $\Omega=\Omega(x,t)$. It can be shown 
\cite{whitham,Lighthill} that $k(x,t)$ and $\Omega(x,t)$ remain constant
for an observer moving at the group velocity $c_g(k)$. Thus, to 
study the asymptotic behavior for large $x$ and $t$ of a wave with given wave 
vector $k$ in deep water, the analysis must be carried out in the frame 
$\bf{R(c_g)}$ traveling  at the velocity
\begin{equation}
c_g(k)=\frac12c(k)=\frac12\sqrt{\frac{g}{k}}
\end{equation}
with respect to the frame $\bf{R}$ where the Euler equations
(\ref{cont}-\ref{BC3}) are written. Therefore,
the linear pattern Ansatz we propose is
\begin{equation}
U(x,z,t)=u\left(k\left(x-\frac12ct\right),t\right)\exp{kz},\quad c=c(k)=\sqrt{\frac{g}{k}}\,.\quad
\label{Udeep}
\end{equation}
%%%%%%%%%%%%%%%%%%%%%%%%%%%%%%%%%%%%%%%%%%%%%%%%%%%%%%%%%%%%%%%%%%%%%%%%%%%%%%
\section{The deep water surface wave problem}
\label{sec: water wave problem}
The first step is to nondimensionalise equations (\ref{cont}-\ref{BC3}). Hence the 
original variables of space, time, velocity components and pressure 
will be normalized 
with $k$, $\sqrt{kg}$, $\sqrt{k/g}$ and $k/\rho g$ respectively.
$k$  is the wave vector of the wave to be described. Hence
the Euler equations for $ -\infty < z 
< \eta $ become
\begin{eqnarray} 
U_{x} + W_{z} &=& 0 \; , \label{cont2}\\
\dot{U} &=& -P^{*}_{x} \;\;  \label{newx2},\\
\dot{W} &=& -P^{*}_{z} -1 \; \label{newz2},
\end{eqnarray}
and the boundary conditions are
\begin{eqnarray}
W &=& 0 \;\; \mbox{for}\;\; z \rightarrow -\infty \label{BC1.2},\\
P^{*} &=& P_0 \;\; \mbox{for} \;\;  z = \eta ,
\label{BC2.2}\\
\eta_t &=& W - U\eta_x \;\; \mbox{for} \;\; 
 z = \eta .\label{BC3.2}
\end{eqnarray}
We integrate the Euler equation in the depth of the fluid. From equation 
(\ref{cont2}) and boundary condition (\ref{BC1.2}) we get
\begin{equation}
W(x,z,t)=-\int^z_{-\infty}U_x(x,\zeta,t)d\zeta\;.
\end{equation}
This expression for $W$  allows us to write  the material derivatives 
$\dot{U}$ and $\dot{W}$ as
\begin{eqnarray}
\dot{U}&=&U_t +UU_x -U_z\int^z_{-\infty}U_xd\zeta\;,\label{Upounto}\\
\dot{W}&=&-\int^z_{-\infty}U_{xt}d\zeta-U\int^z_{-\infty}U_{xx}d\zeta+U_x\int^z_{-\infty}U_{x}d\zeta\;.
\label{Wpunto}
\end{eqnarray}
Using (\ref{Wpunto}) into (\ref{newz2}),  integrating with respect to $z$ and making use of 
(\ref{BC2.2}), we obtain the pressure $P^*(x,z,t)$ as
\begin{eqnarray}\label{intpresion}
P^*&=&\eta - z -\int_{z}^{\eta}d\zeta'\left\{\int_{-\infty}^{\zeta'}U_{xt}d\zeta''
+ U\int^{\zeta'}_{-\infty}U_{xx}d\zeta''
-U_x\int^z_{-\infty}U_{x}d\zeta''\right\}.\nonumber\\
 &+&P_0\end{eqnarray}
Note that  expression (\ref{intpresion}) for the pressure diverges as $z\rightarrow -\infty$ in \textit{Archimedian way}
as it is expected to.
The Newton equation (\ref{newx2}) reads, using (\ref{intpresion}), as
\begin{eqnarray}
U_t+UU_x-U_z\int^{z}_{-\infty}U_xd\zeta'=&-&\eta_x
+\left\{\int^{\eta}_zd\zeta'\int_{-\infty}^{\zeta'}U_{xt}d\zeta''\right.\nonumber\\
 &+&\int^{\eta}_zUd\zeta'\int^{\zeta'}_{-\infty}U_{xx}d\zeta''\nonumber\\
&-&\left.\int^{\eta}_zU_xd\zeta'\int^z_{-\infty}U_{x}d\zeta''\right\}_x\;.
\end{eqnarray}
Now using Leibnitz's rule in the integrals in the right hand side we
obtain
\begin{eqnarray}
U_t+UU_x-U_z\int^{z}_{-\infty}U_xd\zeta'&=&-\eta_x + \int^{\eta}_{z}d\zeta'
\int^{\zeta'}_{-\infty}U_{xxt}d\zeta'' +  \eta_x\int^{\eta}_{-\infty}U_{xt}d\zeta
\nonumber\\
&+&\int^{\eta}_{z}Ud\zeta'\int^{\zeta'}_{-\infty}U_{xxx}d\zeta''
+\eta_xU(\eta)\int^{\eta}_{-\infty}U_{xx}d\zeta \nonumber\\
&-&\int^{\eta}_{z}U_{xx}d\zeta'
\int^{\zeta'}_{-\infty}U_{x}d\zeta'' \nonumber\\
&-&\eta_xU_x(\eta)\int^{\eta}_{-\infty}U_{x}d\zeta \;,\label{newx3}
\end{eqnarray}
with $U(\eta)=U(x,z=\eta,t)$. Finally equation (\ref{BC3.2}) expresses as 
 \begin{equation}
 \eta_{t} + U(\eta)\eta_x +\int^{\eta}_{-\infty}U_xd\zeta=0\;.\label{BC3.3}
 \end{equation}
If the $z$ dependence of $U(x,z,t)$ was known, the surface wave problem in deep
 water could be solved: Eq. (\ref{intpresion}) gives the pressure $P(x,z,t)$, while
 equations (\ref{newx3}) (evaluated in $z= \eta$) and (\ref{BC3.3}) yield a system of coupled nonlinear equations for $U(x,\eta,t)$ and $\eta(x,t)$,
which  completes the solution of the problem. 
However this is not the case because $U(x,z,t)$ is unknown versus $z$.
\section{A Serre-type system of equation}
\label{sec:serre}
The $z$-dependence of $U(x,z,t)$ is the central issue making unsolvable the nonlinear
surface  water wave problem in deep water as  in shallow one. Consequently we use
the deep water Ansatz (\ref{Udeep}),  which reads in dimensionless variables as
\begin{equation}\label{anzatdimen} 
U(\xi,z,t)=u(\xi,t)\exp{z},\quad \mbox{with}\quad \xi= x-\frac12 t\;.
\end{equation}
It implies the change in the differential operators:
\begin{eqnarray}
\frac{\partial}{\partial t } &\rightarrow& -\frac12\frac{\partial}{\partial\xi}
 + \frac{\partial}{\partial t},\nonumber\\
 \frac{\partial}{\partial x} 
&\rightarrow& \frac{\partial}{\partial \xi}\;.
\end{eqnarray}
Then we obtain from (\ref{newx3}) the expression
\begin{eqnarray}\label{expression}
u_t\exp{z}-\frac12 u_{\xi}\exp{z} + \eta_{\xi}&=&-u_{\xi \xi t}\exp{z} + 
(u_{\xi t}\exp{\eta})_{\xi}
+\frac12 u (u_{\xi \xi}\exp{2\eta})_{\xi}\nonumber\\
&-&\frac12 uu_{\xi \xi\xi}\exp{2z}-\frac12 u_{\xi}(u_{\xi}\exp{2\eta})_{\xi}
\nonumber\\
&+&\frac12 u_{\xi}u_{\xi \xi}\exp{2z} - \frac12(u_{\xi \xi}\exp{\eta})_{\xi}
\nonumber\\
&+&\frac12u_{\xi\xi\xi}\exp{z}\label{u(z)}.
\end{eqnarray}
This equation is now taken into account through an averaging over the full depth of the fluid, defined 
by the expression
\begin{equation}\label{zaverage}
\lim_{z_0 \rightarrow -\infty}\frac{1}{\eta-z_0}\int^{\eta}_{z_0}\{*\}d\zeta\;.
\end{equation}
Applying (\ref{zaverage}) to (\ref{expression}) we obtain 
\begin{equation}\label{GNdw1}
\left\{u_{\xi t}\exp(\eta)\right\}_{\xi}-\frac12\left\{u_{\xi \xi}\exp(\eta)\right\}_{\xi}+
\frac{1}{2}\left\{\exp(2\eta)\left[ uu_{\xi\xi}-u_{\xi}^{2}\right]\right\}_{\xi}-\eta_{\xi}=
0.
\end{equation}
Finally, substituting (\ref{anzatdimen}) into (\ref{BC3.3}) we get
\begin{equation}
 \eta_t -\frac12\eta_{\xi}+ (u\exp\eta)_{\xi}=0\label{GNdw2}\;.
\end{equation}
Note that we did not assume that either $\eta(\xi,t)$ or $u(\xi,t)$ vanish as 
$\xi\rightarrow \infty$.

The system  (\ref{GNdw1}) and (\ref{GNdw2}) for $\eta(\xi,t)$ and $u(\xi,t)$ is a 
\textit{$z$-average} of the 
Euler equations resulting from the flow hypothesis (\ref{anzatdimen})
and the subsequent average in the depth (\ref{zaverage}). It is
{\it a deep water analogous} of the Serre (or Green-Naghdi or Su-Gardner) 
system. 
It differs from the result obtained in \cite{leon}. The latter
is a \textit{weighted $z$-integration} of the Euler equations. The crucial
step there was the use of a precise weight in order to regularize 
the Archimedian divergence term present in the pressure. 
In turn 
this produces a \textit{weight-dependent model} the degeneration of which
is eliminated by the requirement that the linear limit possesses phase and group velocities equal to those 
of the deep water system. The present derivation does not need either the weighted average procedure or 
the \textit{ad doc hypothesis} of equality between the linear limits.
Both of them, \textit{$z$-average} and \textit{weighted $z$-integration}, coincide in the 
classical \textit{shallow water} case.

Assuming
that $\eta(\xi,t)$, $u(\xi,t)$ vanish for $\xi\rightarrow \infty$, and setting
$\partial_\theta=\partial_t-\frac12 \partial_\xi$,
 system (\ref{GNdw1}) and (\ref{GNdw2}) can be reduced to the
integro-differential model for $\eta(\xi,t)$~:
\begin{equation}
 \eta_{\theta\theta}+ \eta -\frac{1}{2} \eta^2_\theta-
 \eta_\xi\int^{\xi}_{-\infty}\eta_{\theta\theta}d\xi'-\frac{3}{2}
\eta_{\xi \theta}\int^{\xi}_{-\infty}\eta_{\theta}d\xi'+\frac{1}{2}
\left[ \eta_\xi\left(\int^{\xi}_{-\infty}\eta_{\theta}d\xi'\right)^{2}\right]_\xi =0\;.
\end{equation}
This equation for $\eta(\xi,t)$ is close to classical formulations of model equations
for surface waves in deep water, which are integro-differential models.

\section{An small-aspect-ratio surface wave model}
\label{sec:small aspect}

Up to this point the asymptotic behavior in $\xi$ and $t$ of the surface 
wave profile was not considered. This can be performed by taking
advantage of some small parameter present in the system. Therefore,
 let us rewrite  the
nondimensional function  $\eta$ as
\begin{equation}\label{surface}
\eta(\xi,t) = kaH(\xi,t),
\end{equation}
where the nondimensional function $H(\xi,t)$ represents the surface 
deformation with amplitude $a$ measured from the initial mean water level 
$z=0$, and the product
\begin{equation}\label{stepp}
ka = \epsilon\;,
\end{equation}
is the {\it steepened coefficient} or {\it aspect ratio}, which measures how much the wave profile
is peaked ($\epsilon>1$) or flat ($\epsilon<1$).
The equations (\ref{GNdw1}) and (\ref{GNdw2}) are written explicitly in
terms of $\epsilon$ and $H$  as
\begin{eqnarray}
\left\{\left[u_{\xi t} - \frac12u_{\xi \xi}\right]\exp(\epsilon H)\right\}_\xi
+\left\{\frac{1}{2}\exp(2\epsilon H)\left[ uu_{\xi\xi}-u_{\xi}^{2}\right]\right\}_\xi-\epsilon H_\xi
&=&0,\label{GNdw1bis}\\
\epsilon H_t -\frac12\epsilon H_\xi + \left[u\exp(\epsilon H)\right]_{\xi} &=&0.
\label{GNdw2bis}
\end{eqnarray}

In fact, (\ref{GNdw1bis}) and (\ref{GNdw2bis}) are valid for 
any orders in $\epsilon$, but they are dispersive and 
strongly nonlinear, so untractable analytically.
Nevertheless, a perturbative theory can be carried out in terms of 
$\epsilon$  by considering flat propagating waves, i.e., $\epsilon<1$. 
The linearized system yields the nondimensional form of 
the  dispersion relation
\begin{equation}\label{nondiomega}
\Omega = \frac{1}{2}.
\end{equation}
Nonlinearity causes
deviations from the latter (Stokes' hypothesis) which can be taken into
account through
\begin{equation}\label{omegapertur}
\Omega = \frac{1}{2} + \epsilon + \epsilon^2 + ...\,\,.
\end{equation}
Hence the phase becomes
\begin{equation}
\xi -\Omega t = \xi -\frac{1}{2} t- \epsilon t - \epsilon^2 t +...\,,
\end{equation}
and we can define new variables $y$, $\tau$, $\nu$, ..., as
\begin{equation}\label{varlentes} 
y = \xi - \frac{1}{2} t, \quad \tau = \epsilon t, \quad \nu = \epsilon^2 t, ...\,.
\end{equation}
The function $ H(\xi,t) $ must then be regarded as a function of the new
independent variables $y$, $\tau$, $\nu$, ... \cite{sandri, pereira}, i.e., as
$H(y, \tau, \nu, ...,)$. On the other
hand, the  derivation operators with respect to $x$ and $t$ become
\begin{equation}\label{operators} 
\frac\partial{\partial\xi} =\frac\partial{\partial y}, \quad \quad
\frac\partial{\partial t} = - \frac{1}{2}\frac\partial{\partial y} +
\epsilon \frac\partial{\partial\tau} +\epsilon^2 \frac\partial{\partial\nu}, ...\;.
\end{equation}
Since $\exp(\epsilon H) = 1 + \epsilon H + O(\epsilon^2)$, we can retain in
equations (\ref{GNdw1bis}) and  (\ref{GNdw2bis}), after the leading order $\epsilon^0$, 
\textit{both}  orders  $\epsilon^0$ and $\epsilon$,  and so on. 
Thus, Eq. (\ref{GNdw2bis}) reads as
\begin{equation}
\left(-\frac{1}{2}\partial_y + \epsilon \partial_{\tau} +O\left(\epsilon^2\right)\right)\epsilon H 
-\frac{1}{2}\partial_y\epsilon H + \left[u\left(1+\epsilon H +O\left(\epsilon^2\right)\right)\right]_y=0\;.
\end{equation}
This equation gives $u$ in terms of derivatives and antiderivatives of $H$, as
\begin{equation}
 u=\epsilon H -\epsilon^2\left(H^2 + \int^y_{-\infty}H_{\tau}dy'\right) + O\left(\epsilon^3\right)\;,
\end{equation}
where was not assumed that $u$ or $H$ (and derivatives) vanish as $y \rightarrow -\infty$. 
Instead of such assumption,
the above equation implies the relation
\begin{equation}
u(-\infty,\tau) = \epsilon H(-\infty,\tau)\left[1 - \epsilon H
(-\infty,\tau) + O\left(\epsilon^2\right)\right]. 
\end{equation}
Reporting the expression of $u$ into (\ref{GNdw1bis}) and retaining terms in $\epsilon$ and $\epsilon^2 $, we
obtain the equation
\begin{equation}
 -2\epsilon^2 H_{\tau yy} + \epsilon H_y +\epsilon H_{yyy}
 - \frac{3}{2}\epsilon^2 (HH_y)_{yy}=0\;,
\end{equation}
which in terms of the dimensionless variables $\xi, t$ and $ \eta(\xi,t)$ reads as
\begin{equation}\label{eqeta}
2\eta_{\xi\xi t} = \eta_\xi - \frac{3}{2}(\eta\eta_\xi)_{\xi\xi}\;,
\end{equation}
or in terms of variables with dimension:
\begin{eqnarray}\label{eqetadim}
2\sqrt{\frac{k}{g}}\eta_{\xi\xi t} = k^2\eta_\xi- \frac{3}{2}k(\eta\eta_\xi)_{\xi\xi}\,.
\end{eqnarray}
This equation is the main result of our work. It describes the asymptotic nonlinear and
dispersive evolution of small-aspect-ratio waves of a Fourier wave vector $k$ in deep water.
It has $k$-dependent coefficients. Equation 
(\ref{eqetadim}) can be considered  as belonging to both of the two categories: that of KdV models 
(KdV, modified KdV, Benjamin-Bona-Mahony-Peregrine, Camassa-Holm, etc.) describing evolutions of wave profiles and that of  NLS-type 
equations (modified NLS \cite{kortewegmodel}, Davey-Stewartson \cite{davey}, etc.) describing modulation of wave profiles and having
$k$-dependent coefficients.

%--------------------------------------------
\section{Mathematical properties}
\label{sec:Mathematical}
\subsection{Lax pair}
\label{sec:Lax} 
In this section we assume that $\eta$ and its derivatives vanish as 
$\xi \rightarrow \pm\infty$.
Equation (\ref{eqeta}) has the dispersion relation 
\begin{equation}\label{Omega}
 \Omega=\frac12\;,
\end{equation}
as required,  because (\ref{eqeta}) was derived in a frame moving 
at the group velocity associated with the Euler equations in deep water.
Making use of the scalings 
\begin{equation}
 \eta(\xi,t) \longrightarrow \frac{2}{3}\eta(\xi,t)\;,\quad 2\frac\partial{\partial t} \longrightarrow \frac\partial{\partial t}\;,
\end{equation}
 Eq.(\ref{eqeta}) becomes
\begin{equation}\label{eqetascaling}
\eta_{\xi\xi t} = \eta_\xi - (\eta\eta_\xi)_{\xi\xi},
\end{equation}
and integrating once yields  
\begin{equation}\label{integrat}
 \eta_{\xi t}=\eta-(\eta\eta_\xi)_\xi\;.
\end{equation}
Let us consider the function $F$ defined by
\begin{equation}\label{fonctionF}
 F^3=1-3\eta_{\xi\xi}\;,
\end{equation}
or in terms of $\phi=\eta_{\xi\xi}$, by
\begin{equation}
 F^3=1-3\phi.
\end{equation}
One of the most remarkable properties of F is that  it allows us to build non-trivial 
conserved quantities. The two first ones are
\begin{eqnarray}\label{consquant}
F_t &=-&(\eta F)_\xi\;, \nonumber\\
\left(2\phi_{\xi}^2F^{-7}\right)_t &= & 
\left[\left(1+\phi\right)F^{-1}-2\eta\phi_{\xi}^2F^{-7}\right]_\xi.
\end{eqnarray}
The Lax pair also can be constructed by means of $F$ and reads
\begin{eqnarray}
 \hat{L}&=&\frac{\partial}{\partial \xi}+i\lambda F \hat{\sigma_3}+
 \frac{1}{2}\frac{\eta_{\xi\xi\xi}}{F^3}\hat{\sigma_1},\nonumber\\
 \hat{M}&=&-\frac{1}{2}\frac{\eta\eta_{\xi\xi\xi}}{F^3}\hat{\sigma_1} -
i\lambda \eta F \hat{\sigma_3}-\frac{i}{8\lambda}\frac{1-\eta_{\xi\xi}}{F}\hat{\sigma_3}+
\frac{1}{4}\frac{\eta_{\xi\xi}}{\lambda F}\hat{\sigma_2},
\end{eqnarray}
where $ \hat{\sigma_1},\hat{\sigma_2}$ and $\hat{\sigma_3} $ are the Pauli matrices, $\lambda$
the spectral parameter and Eq. (\ref{integrat}) is obtained as the 
classical Lax equation $ \hat{L}_t=[\hat{L},\hat{M}]$.
\subsection{Reduction to the Bullough-Dodd equation}
\label{sec:Bullough-Dodd equation}
Through the change of variables from $\xi$ to $p$ and the change of functions 
from $\eta(\xi,t)$ to $r(p,t)$ defined by
\begin{eqnarray}
p &=& \int^\xi Fd\xi \;,\label{vary2}\\
1-3\eta_{\xi\xi}&=&\exp(r) \;,\label{funcr2}
 \end{eqnarray}
it is found that $r(p,t)$ satisfies the Bullough-Dodd equation (see Appendix)
\begin{equation}\label{Bullough-Dodd}
r_{pt}=\frac{1}{9}\left\{ \exp\left(\frac{2}{3}r\right) -\exp\left(-\frac{1}{3}r\right)\right\}\;.
\end{equation}
The Bullough-Dodd equation was introduced in \cite{DoddBullough}, in which 
some non-trivial conserved densities  were also  exhibited.
 Its Lax pair and  complete integrability were shown in \cite{mikhailov}. 
 The change of variable and of function (\ref{vary2}) and  (\ref{funcr2}) 
 leading from (\ref{integrat}) to (\ref{Bullough-Dodd}) were introduced 
 in the context of a systematic study of short-wave dynamics in long-wave 
 model equations in  references \cite{manna1,manna2,manna3,fakir}.
%%%%%%%%%%%%%%%%%%%%%%%%%%%%%%%%%%%%%%%%%%%%%%%%%%%%%%%%%%%%%%%%%%%%%%%%%%%%%%%%%%%%%%%%%%%%
\subsection{Symmetries}
\label{sec:sym}
Equation (\ref{eqetascaling}) is invariant under the discrete transformation
\begin{equation}
 \eta(\xi,t) \rightarrow \eta(-\xi,-t)\;,
\end{equation}
and can be written as the conservation law
\begin{equation}
 \partial_t (\eta_{\xi\xi})= \partial_\xi(\eta-(\eta \eta_\xi)_\xi)\;.
\end{equation}
It is invariant under the Galilean group of transformations 
\begin{equation}\label{Galileo}
t'=t\;, \quad \quad x'=\xi + V t\;,\quad \quad \eta(\xi,t)=\eta_0 + S(\xi,t),
\end{equation} 
in which $V$ and $\eta_0$ are constants with $\eta_0=-V$. In this case indeed, 
 Eq. (\ref{eqetascaling}) transforms  into
\begin{equation}\label{gal}
S_{x'x't'} = S_{x'}- (SS_{x'})_{x'x'} \;.
\end{equation}
Consequently, (\ref{eqetascaling}) must be seen as a member of a \textit{family} 
of equations parametrized by the speed $V$
of the Galilean group of transformations (\ref{Galileo}). This invariance
comes  from the invariance under (\ref{Galileo}) of the Euler equations. Last but not the 
least, the equation (\ref{eqetascaling}) - or (\ref{integrat}) -  has the  Lorentz invariance 
\begin{equation}
\xi \rightarrow \kappa \xi, \quad t \rightarrow \frac{t}{\kappa}, 
\quad \eta \rightarrow \kappa^2\eta, \end{equation}
with $\kappa$ a real arbitrary parameter,
which is undoubtedly related to the integrability of the associated  Bullough-Dodd equation.
%%%%%%%%%%%%%%%%%%%%%%%%%%%%%%%%%%%%%%%%%%%%%%%%%%%%%%%%%%%%%%%%%
\section{Periodic wave and  limiting wave}
\label{sec:Stokes}
\subsection{Harmonic solution and perturbative approach}

A periodic wave solution of (\ref{eqetadim}) is predicted by linear analysis.
 It is sinusoidal, of small amplitude and travels with phase velocity $(1/2)\sqrt{g/k}$. The question is: up to what
values of its amplitude and velocity does the periodic wave  exist? 
This is still possible for wave amplitudes large enough so that higher terms in the  perturbation series can 
no longer be neglected in Eq. (\ref{eqetadim}).
Then the  wave is not  exactly sinusoidal,  its velocity is not 
exactly $\frac12\sqrt{g/k}$, and the
 \textit{limiting wave} has the greatest height before breaking. 
In order to find this \textit{steep rotational Stokes wave}, let us start with 
equation (\ref{eqetadim}) in the frame {\bf R} with coordinates $(x, t)$, which reads as
\begin{eqnarray}\label{eqetadim2}
2\sqrt{\frac{k}{g}}\eta_{xxt} = k^2\eta_x- \eta_{xxx}  - \frac{9}{2}k\eta_{x}\eta_{xx}
 -\frac{3}{2}k\eta\eta_{xxx}\,.
\end{eqnarray}
The radius of curvature is defined by
\begin{equation}\label{radius}
R=\frac{\lbrace1+ (\eta_{x})^2\rbrace^{3/2}}{\eta_{xx} }\,.
 \end{equation}
Hence Eq. (\ref{eqetadim2}), within the allowed order, can be written as 
\begin{equation}\label{eqR}
-2\sqrt{\frac{k}{g}}\frac{{R_t}}{R^2}=k^2\eta_x+\frac{{R_x}}{R^2}-\frac{9}{2}k\frac{{\eta_x}}
{R}+\frac{3}{2}k\eta\frac{{R_x}}{R^2}\,.
 \end{equation}
We assume a progressive periodic wave in the variable $z=x-ct$, hence we have
\begin{equation}\label{eqR(z)}
2c\sqrt{\frac{k}{g}}R_z=k^2\eta_zR^2+R_z-\frac{9}{2}k\eta_zR+\frac{3}{2}k\eta R_z\,.
 \end{equation}
Let $\eta$ be the height of the wave, i.e.,
\begin{equation}\label{hmax}
\eta=\eta_{crest}-\eta_{trough},
\end{equation}
and assume that the wave approaches the limiting wave of height $\eta = \eta_{max} $
and that this value is reached for $z=z_0$; i.e, $\eta(z_0)=\eta_{max}$ 
with $z_0=x_0-c_Lt_0$, where $c_L$ is the  phase velocity at the limit. At the limit, on the one hand, the radius of 
curvature $R(z_0)$ is zero \cite{longuet1,longuet2}, and  on the other hand its derivative $R_z(z_0)$ is not well 
defined. Nevertheless, equation (\ref{eqR(z)}) can be satisfied if 
\begin{equation}\label{cesolution1}
2c_L\sqrt{\frac{k}{g}}=1+\frac{3}{2}k\eta_{max}\,.
\end{equation}
To determine the limit $k\eta_{max}$ and 
$c_L$,  one further equation is required. This equation will come from the Stokes series for 
the periodic solution of small amplitude of Eq. (\ref{eqetadim2}). Hence we consider a periodic solution
\begin{equation}\label{etaexpansion}
 \eta(z) =\delta\lbrace \eta_0(z) + \delta \eta_1(z) +\delta^2 \eta_2(z)+...\rbrace\,,
\end{equation}
with $\delta$ a small amplitude defined as
\begin{equation}\label{epsilondefi}
\delta=\frac12(\eta_{crest}-\eta_{trough})=\frac12 \eta.
\end{equation}
To avoid secular terms, the velocity  $c$ must  be also expanded as
\begin{equation}\label{cexpansion}
c=c_0+\delta c_1+\delta^2c_2 +...\,.
\end{equation}
We write Eq. (\ref{eqetadim2}) in terms of $z$, report expansions 
(\ref{etaexpansion}) and (\ref{cexpansion}) into it, and then isolate order by 
order 
the coefficients of each power of $\delta$. We  obtain this way  a series of equations, whose leading order $\delta$ is
\begin{equation}\label{order1}
\left(1-2c_0\sqrt{\frac{k}{g}}\right)\eta_{0,3z}-k^2\eta_{0,z}=0\,.
 \end{equation}
An even solution of (\ref{order1}) is
\begin{equation}
 \eta_0(z)=\cos(kz) \quad \mbox{with}\quad c_0=\sqrt{\frac{g}{k}}\,.\label{solorder1}
\end{equation}
Using the solution (\ref{solorder1}), the equation at order $\delta^2$ is 
\begin{equation}
 \eta_{1,3z}+k^2\eta_{1,z}=-2c_1\sqrt{\frac{k}{g}} k^3\sin(kz)+3k^2\sin(2kz)\,.
\end{equation}
The trouble is that the term in $\sin(kz)$ resonates with the operator on the left-hand-side. This is  a \textit{secular term}. It can be  eliminated 
by choosing $c_1=0$. Thus the solution
at this order is
\begin{equation}
 \eta_1(z)=\frac{k}{2}\cos(2kz) \quad \mbox{with}\quad c_1=0\,.
\end{equation}
At order $\delta^3$ we find
\begin{equation}
  \eta_{2,3z}+k^2\eta_{2,z}=-\left(2c_2\sqrt{\frac{k}{g}} k^3-\frac{3k^5}{8}\right)\sin(kz)-
\frac{81k^5}{8}\sin(3kz)\,,
\end{equation}
and the solution free of secularity is
\begin{equation}
 \eta_2(z)=\frac{27k^2}{64}\cos(3kz)\quad \mbox{with}\quad c_2=\frac{3k^2}{16}\sqrt{\frac{g}{k}}\,.
\end{equation}
Thus, the solution $\eta(z)$ at order $\delta^3$ is
\begin{eqnarray}\label{soleta}
 \eta(z)&=&\delta \cos(kz)+ \delta^2\frac{k}{2}\cos(2kz)+\delta^3\frac{27k^2}{32}\cos(3kz)+...,\nonumber\\
\mbox{with } c&=&\sqrt{\frac{g}{k}}\left(1 + \delta^2\frac{3k^2}{16}+...\right)\,.
\end{eqnarray}
Now, when the solution (\ref{soleta}) approaches the limiting wave solution we must 
have 
\begin{equation}
\delta \longrightarrow \frac{\eta_{max}}{2}\quad
 \mbox{and}\quad c\longrightarrow c_L ,
\end{equation}
hence
\begin{equation}\label{cesolution2}
c_L\sqrt{\frac{k}{g}}=1+\frac{3(k\eta_{max})^2}{64}\,.
\end{equation}
This is the second equation we were looking for. From (\ref{cesolution1}) and 
(\ref{cesolution2}) we obtain
\begin{equation}\label{serieresults}
k\eta_{max}=0.697, \quad\quad\quad\quad c_L\sqrt{\frac{k}{g}}=1.022\,.
 \end{equation}
In the irrotational case the classical result found a long time ago
 by J. H. Michell \cite{Michell} for $k\eta_{max}$ and  recent
computations  for $c_L$ \cite{Cokelet} yield the values
\begin{equation}\label{MicCok}
k\eta_{max}=0.892,\quad\quad\quad c_L\sqrt{\frac{k}{g}}=1.092.
\end{equation}
Comparing (\ref{serieresults}) and  (\ref{MicCok}) we find the percentage 
errors (100\% times the relative errors)
\begin{equation}\label{errors}
(k\eta_{max})_{error}=-21\%,\quad\quad\quad \left(c_L\sqrt{\frac{k}{g}}\right)_{error}=
-6,4\%
\end{equation}
Deviations of $k\eta_{max}$ and  $c_L$, calculated from (\ref{eqetadim2}) in
relation of (\ref{MicCok}) are due to three mains factors:
\begin{itemize}
\item the results (\ref{serieresults}) come from a computation at
second order in $\delta$ only, while  (\ref{MicCok}) were obtained using 
theories at least at fifth order in  $\delta$,
\item the classical results were obtained from the complete Euler equations,
\item our fluid is rotational.
\end{itemize}
%%%%%%%%%%%%%%%%%%%%%%%%%%%%%%%%%%%%%%%%%%%%%%%%%%%%%%%%%%%%%%%%%%%%%%%%%%%%%%%%
\subsection{Exact analytical approach to the limiting wave} 
\label{sec:numerical}
We start with the equation (\ref{eqeta}) in the referential {\bf R} relative to the
 variables ($\xi,t$),  i.e.,
\begin{equation}\label{eqetaint2}
 2\eta_{\xi t}=\eta-\eta_{\xi\xi}-\frac34(\eta^2)_{\xi\xi}\;.
\end{equation}
 We look for a traveling wave with speed $c$, 
\begin{equation}
 \eta=\eta(z) \;,\quad z=k(\xi-ct)\;,
\end{equation}
so that (\ref{eqetaint2}) can be written as
\begin{equation}
\left [\frac{3}{4}\eta^2 + a\eta\right]_{zz}=\eta,
\end{equation}
where we have set $a=1-2c$.
Now multiplying both sides by 
$$
\left[\frac{3}{4}\eta^2 + a\eta\right]_{z}\;,
$$
and integrating once, we obtain
\begin{equation}
 \left[\left(\frac34\eta^2+a\eta\right)_z\right]^2=
\eta^3+a\eta^2+K,
\end{equation}
$K$ being some integration constant.
And hence
\begin{equation}
dz=\pm\frac{3\eta+2a}{2\sqrt{\eta^3+a\eta^2+K}}d\eta.
\label{dzdeta}
\end{equation}
Setting $\eta=aY$, $ z=Z\sqrt{|a|}$ and $K=|a|^3p$
transforms (\ref{dzdeta}) into
\begin{equation}
dZ=\pm\frac{3Y+2}{2\sqrt{p+\varepsilon\left(Y^3+Y^2\right)}}dY,
\label{dzdy}
\end{equation}
$\varepsilon$ being the sign of $a$.
Then we look for some regular periodic solution. It attains its minimal value $Y_1$ for some $Z$,
say $Z=Z_1$, and then will grow up to some maximal value $Y_2$, reached at $Z=Z_2$ for the first time after $Z_1$. $dY/dZ $ must be well-defined, real and positive for all $Z$ between $Z_1$ and $Z_2$
and all $Y$ between $Y_1$ and $Y_2$. Further $dY/dZ $ must be zero for $Z=Z_1$ and $Z_2$. 
The conditions $Z_1<Z_2$ and $Y_1<Y_2$ must also be satisfied.

\begin{figure}
\begin{center}
\includegraphics[width=7cm]{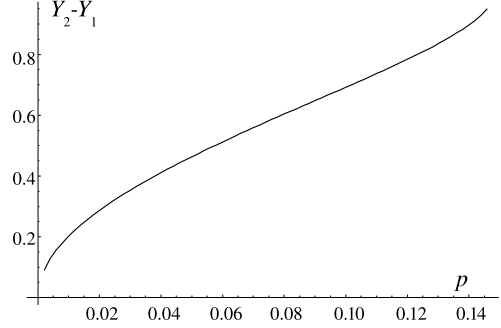}
\caption{ The normalized amplitude $Y_2-Y_1$ of the Stokes wave vs the parameter $p$.}
\label{ampvsp}
\end{center}
\end{figure}

\begin{figure}
\begin{center}
\includegraphics[width=7cm]{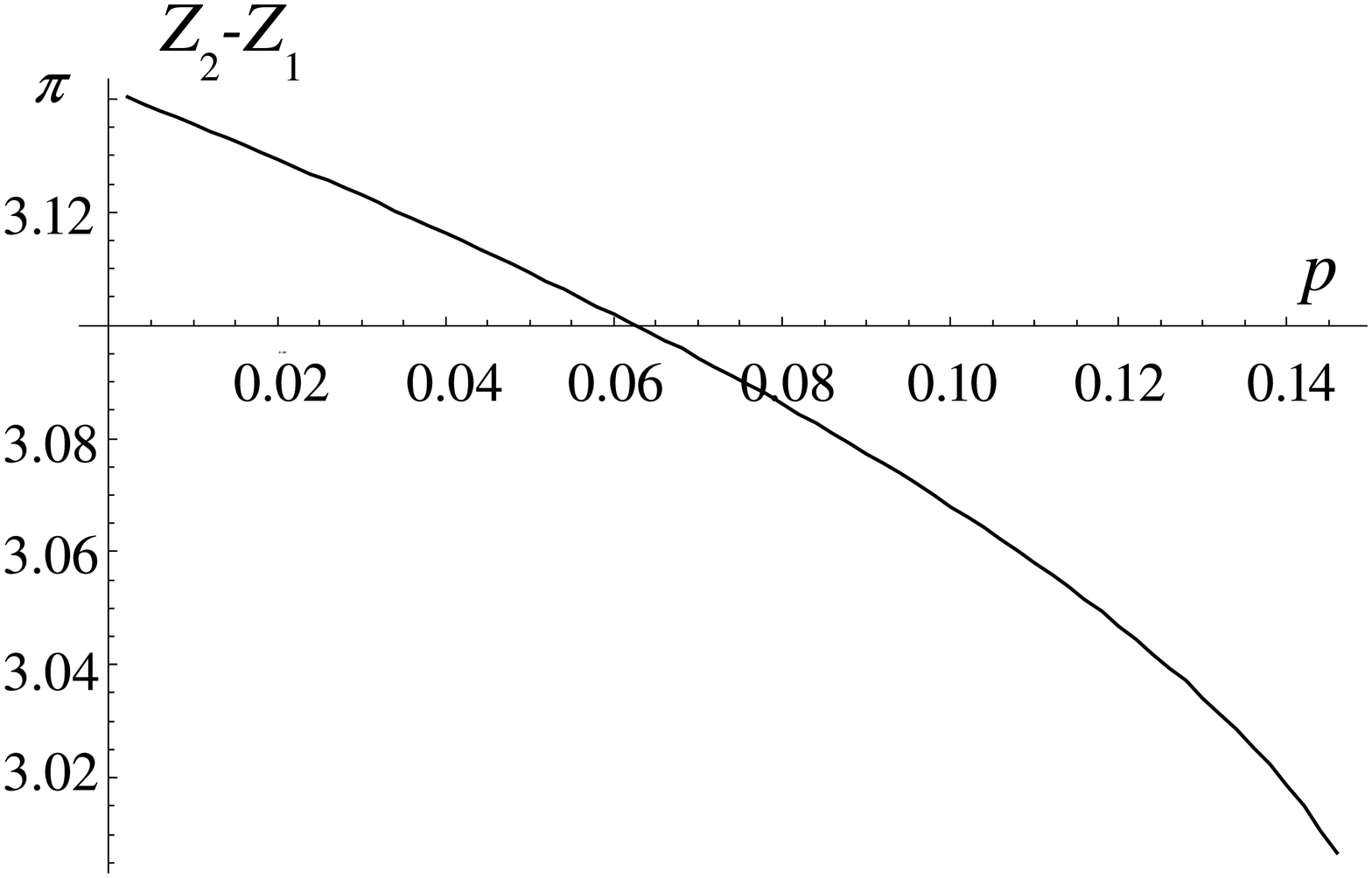}
\caption{  The normalized wavelength $Z_2-Z_1$ of the Stokes wave vs the parameter $p$.}
\label{pervsp}
\end{center}
\end{figure}

\begin{figure}
\begin{center}
\includegraphics[width=7cm]{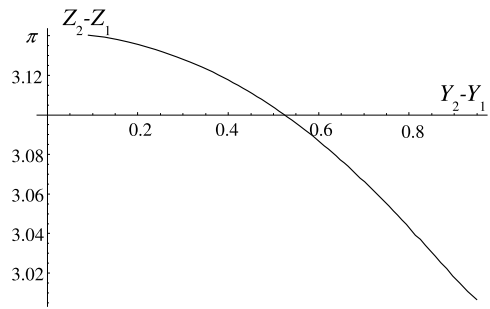}
\caption{ The normalized wavelength $Z_2-Z_1$  of the Stokes wave vs its normalized amplitude $Y_2-Y_1$.}
\label{pervsamp}
\end{center}
\end{figure}

$f(Y)=Y^3+Y^2$ presents two local extrema, $f(-2/3)=4/27$ and $f(0) =0$. Hence $ p+\varepsilon f(Y) $
has more than one zero if $p$ is between 0 and $-\varepsilon 4/27$.
Since $dZ/dY$ is zero for $Y=-2/3$, and $dY/dZ$ must remain finite one the whole interval, 
the value $Y=-2/3$ is excluded, and $Y_1$ and $Y_2$ must be the two largest zeros of
$ p+\varepsilon f(Y)$, with $Y_1<0<Y_2$.
For $\varepsilon =+1$, $ p+\varepsilon f(Y)$ is negative in this interval. Hence, since its square root must be defined, the case 
$\varepsilon =1$ is excluded.
The period of this solution, which is the wavelength of the Stokes wave in normalized units,
is
\begin{equation}
Z_2-Z_1=\int_{Y_1}^{Y_2}\frac{3Y+2}{2\sqrt{p-Y^2-Y^3}}dY.
\label{period}
\end{equation}
The amplitude $Y_2-Y_1$ is given by the solution of the cubic polynomial equation $Y^3+Y^2-p=0$, 
it is plotted vs $p$ in Fig. \ref{ampvsp}.
 The integral (\ref{period}) can be computed 
numerically, which yields the normalized wavelength vs the integration parameter $p$ as shown on Fig. \ref{pervsp}.
Finally the wavelength is shown versus the amplitude in Fig. \ref{pervsamp}.
The maximum amplitude it attained with the minimal wavelength for $p=4/27$.
Then the quantity $p-Y^2-Y^3$ factorizes explicitly, and after simplification
we get 
\begin{equation}
\frac{dZ}{dY}=\frac{3\sqrt3}2\frac1{\sqrt {1-3Y}},
\end{equation}
which is straightforwardly integrated, and then inverted, to yield
\begin{equation}
Y=\frac 13-\frac{Z^2}9,
\label{limw}
\end{equation}
in which we have set arbitrarily $Z_1=3$ (translation invariance). Formula (\ref{limw}) is completed using
parity and periodicity to
\begin{equation}
Y=\frac 13-\frac19\left(Z-6E\left(\frac{Z+3}6\right)\right)^2,
\label{limw2}
\end{equation}
$E$ denoting the integer part. 
Coming back to the original variables,
we obtain the expression of the limiting Stokes wave, as 
\begin{equation}
\eta=\frac{\lambda^2}{36}\left[\left(\frac{2z}\lambda-2 E\left(\frac z\lambda+\frac12\right)\right)^2
-\frac13\right],
\label{limeta}
\end{equation}
in which $\lambda=6\sqrt{-a}$ is the wavelength (it has been seen that
 $\varepsilon=\mbox{sign}\,( a)=-1$).
The velocity is 
\begin{equation}
c=\frac12-\frac{a}2=\frac12+\frac{\lambda^2}{72}. 
\end{equation}
The total amplitude is $-a={\lambda^2}/{36}$.

Coming back to the dimensioned variables, we obtain
\begin{equation}
\eta=\frac{k{\lambda'}^2}{36}\left[\left(\frac{2\zeta}{\lambda'}-2 E\left(\frac \zeta{\lambda'}
+\frac12\right)\right)^2
-\frac13\right],
\label{limdim1}
\end{equation}
in which $\lambda'=\lambda/k$ is the dimensioned wavelength, and
\begin{equation}
\zeta=x-\frac{k^{3/2}\lambda^2}{72}\sqrt{g}t.
\end{equation}
We need to identify $k$. The exact linear wave must be periodic with wavelength  $\lambda_l=2\pi/k$. 
Assuming the same value of $k$ for the same wavelength in the nonlinear case 
($\lambda'=\lambda_l$), we get (dropping the prime)  
\begin{equation}
\eta=\frac{2\pi \lambda}{36}\left[\left(\frac{2\zeta}{\lambda}-2 E\left(\frac \zeta{\lambda}
+\frac12\right)\right)^2
-\frac13\right],
\label{limdim}
\end{equation}
and $\zeta=x-c_Lt$, with 
\begin{equation}\label{limgraf}
c_L\sqrt{\frac kg}=\left(\frac12+\frac{\pi^{2}}{18}\right)\simeq 1.0483.
\end{equation}

\begin{figure}
\begin{center}
\includegraphics[width=7cm]{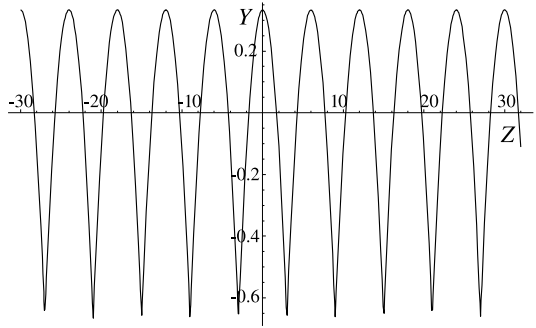}
\caption{The limiting Stokes waves in dimensionless form. }
\label{figstkw}
\end{center}
\end{figure}

\begin{figure}
\begin{center}
\includegraphics[width=7cm]{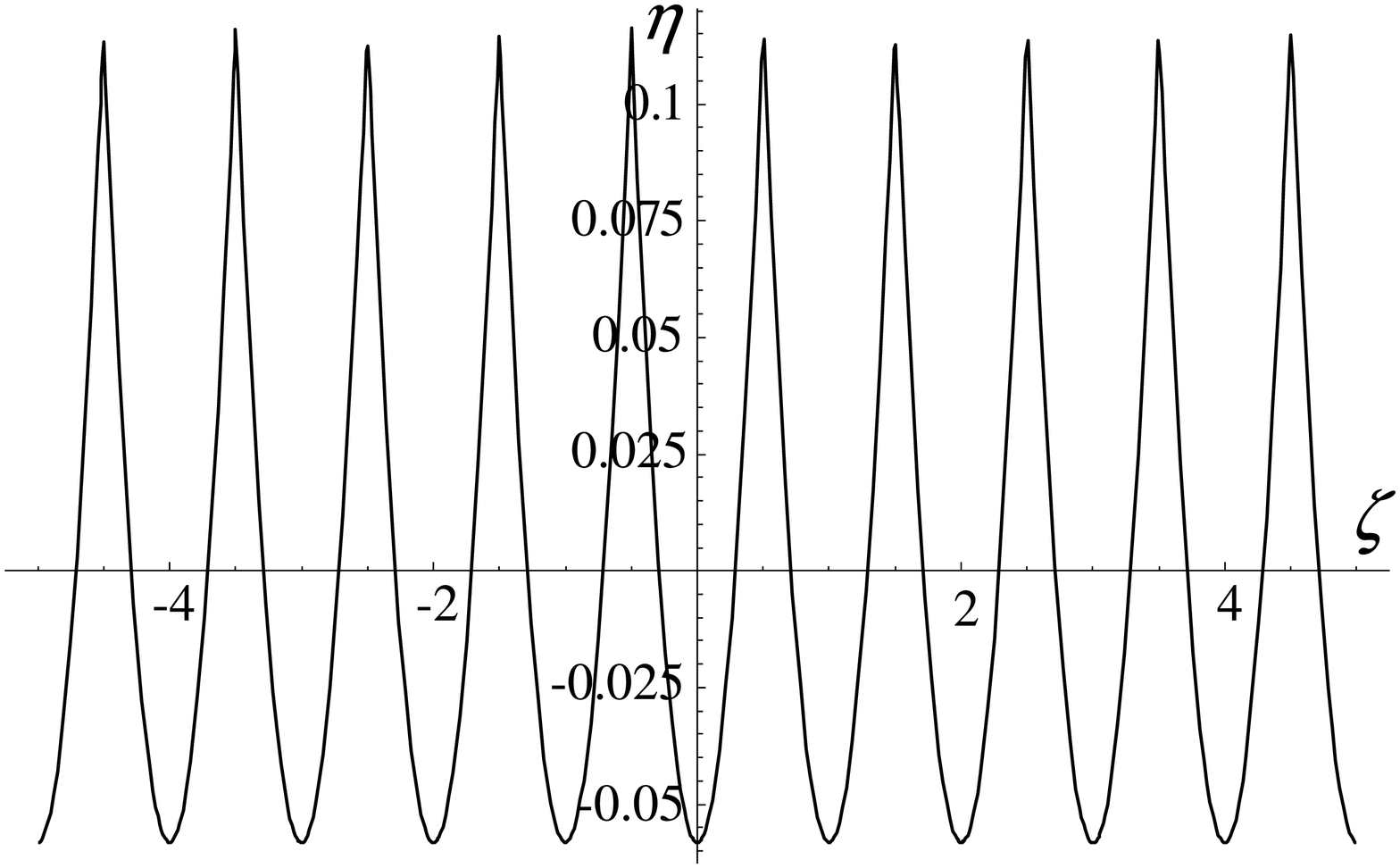}
\caption{The limiting Stokes waves in dimensioned form, for $\lambda=1$.  }
\label{stkdim}
\end{center}
\end{figure}

The limiting Stokes waves, in the dimensionless form (\ref{limw2}), is plotted on Fig.~\ref{figstkw}.
Its dimensioned form (\ref{limdim}) is plotted on Fig. \ref{stkdim} 
for $\lambda=1$. 
Recall that $Y/\eta$ is negative!
Now the percentage errors using (\ref{limdim}) and (\ref{limgraf}) are
\begin{equation}\label{errors2}
(k\eta_{max})_{error}=+22\%,\quad\quad\quad \left(c_L\sqrt{\frac{k}{g}}\right)_{error}=
-4\%.
\end{equation}
\subsection{The evolution of the limiting Stokes wave}
\begin{figure}
\begin{center}
\includegraphics[width=7cm]{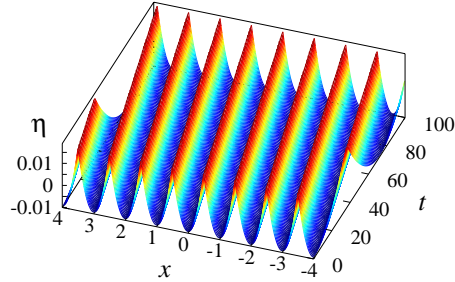}
\caption{The evolution of the limiting Stokes wave, as  computed numerically. }
\label{stoknum}
\end{center}
\end{figure}

The evolution of the limiting Stokes wave is checked numerically.
We use the Eq. (\ref{eqetaint2}), written in frame traveling at velocity 1, as
\begin{equation}\label{eqetaint3}
 2\eta_{\xi t}=\eta-\frac34(\eta^2)_{\xi\xi}\;.
\end{equation}
 Eq. (\ref{eqetaint3})
 is integrated with respect to $x$, then we apply the Fourier transform defined as
\begin{equation}
 y(\xi,t)=\int\hat y(\sigma,t)e^{2i\pi\sigma \xi}d\xi,
\end{equation}
to get 
\begin{equation}\label{eqfou}
 \hat \eta_{t}=\frac 1{4i\pi\sigma}\hat\eta-\frac34 i\pi\sigma\widehat{\eta^2}\;.
\end{equation}
Then the time evolution of $\hat\eta$ is computed by means of a standard 4th order Runge-Kutta algorithm.
The Fourier transforms are computed using a standard fast Fourier transform algorithm.
It must be noticed that the term representing the antiderivative in Eq. (\ref{eqfou})
 is not defined for $\sigma=0$. The corresponding term in the discrete scheme is set to zero, 
which assumes a zero mean value. The mean value of the limiting Stokes wave (\ref{limdim}) is easily 
computed, it is zero as required.
Further, the fact that (\ref{limdim}) is not continuously derivable induces a high-frequency numerical
instability. The latter is removed using a spectral filter in the numerical scheme, which attenuates the highest frequencies. It is checked that this filter does not affect the spectrum of the 
limiting Stokes wave itself. 
The result of the computation is plotted on Fig. \ref{stoknum}.
%%%%%%%%%%%%%%%%%%%%%%%%%%%%%%%%%%%%%%%%%%%%%%%%%%%%%%%%%%%%%%%%%%%%%%%%%%%%%%%%
\section{Conclusions and final comments}
We have obtained a {\it deep water equivalent} of the widely known Serre 
shallow water system of equations. This was done via an appropriated Ansatz for
the vertical coordinate dependence of the velocities. This procedure allowed us
for a description of the nonlinear and dispersive dynamics of the wave profile %with wave vector $k$
itself, avoiding the classical route of wave modulation of a short wave train.    
From the system, we were able to derive an asymptotic model equation (in times 
and space) for {\it short waves}. The new model is the short wave analogous of 
KdV in surface waves and, as KdV, shown complete integrability.
The progressive Stokes wave was studied analytically and numerically. The
results for the limiting wave height and the limiting phase velocity, within the degree of 
approximation of the model, are in accordance with classical results.
Important future works and open issues to be study are :  a) the analysis 
of the Stokes limiting wave of the system for the free surface elevation and 
the free surface velocity, b) inclusion of surface tension, small viscosity
as well as higher spatial dimension, c) the achievement of asymptotic systems of 
 higher degree than two in the aspect-ratio parameter,
d) is the new asymptotic model the first term of a hierarchy of 
integrable models, as KdV is?
e) last, but not the least, an important open issue is: can we
go further in the unexpected bridge established by the system between 
nonlinear and dispersive short wave dynamics in 
fluid mechanical and two-dimensional integrable relativistic field theory? 
\label{sec:conclusion}
%%%%%%%%%%%%%%%%%%%%%%%%%%%%%%%%%%%%%%%%%%%%%%%%%%%%%%%%%%%%%%%%%%%%%%%%%%%%%%%
\section*{Acknowledgments}
M. A. M wish to thank the IFT-UNESP for their hospitality and 
FAPESP (Brasil) for financial support.
\section*{Appendix A. Reduction to the Bullough-Dodd equation}
The change of variables from $\xi$ to $p$ and the change of functions 
from $\eta(\xi,t)$ to $r(p,t)$ 
\begin{eqnarray}
p &=& \int^\xi Fd\xi, \\ F&=&(1-3u_{\xi\xi})^{\frac13},\label{vary3}\\
1-3\eta_{\xi\xi}(\xi,t)&=&\exp(r(y,t)) \;,\label{funcr}
\end{eqnarray}
together with the conservation of $F$
\begin{equation}
F_t =-(\eta F)_\xi,
\end{equation}
gives for any function $M(\xi,t)$ the relations
\begin{eqnarray}
\frac{\partial M(\xi,t)}{\partial \xi}&=&F\frac{\partial M(p,t)}{\partial
p},\label{relationA}\\
\frac{\partial M(\xi,t)}{\partial t}&=&-(\eta F)\frac{\partial M(p,t)}{\partial p}+
\frac{\partial M(p,t)}{\partial t}.\label{relationB}
\end{eqnarray}
From (\ref{relationB}) follows that
\begin{equation}\label{relationC}
\frac{\partial^2 M(\xi,t)}{\partial t \partial \xi}+\eta 
\frac{\partial^2 M(\xi,t)}{\partial \xi^2}=-\eta_{\xi}F
\frac{\partial M(p,t)}{\partial p}+F\frac{\partial^2 M(p,t)}{\partial p
\partial t}.
\end{equation}
We take the derivative of the evolution equation (\ref{integrat}) with respect to $\xi$,  once and then twice.
Using the definition of $F$, it  yields
\begin{eqnarray}
\eta_{\xi\xi t}+\eta \eta_{\xi\xi\xi}&=&\eta_{\xi}F^3,\label{once}\\
\eta_{\xi\xi\xi t}+\eta \eta_{\xi\xi\xi\xi}&=&\eta_{\xi \xi}F^3 -4\eta_{\xi}F
\eta_{\xi\xi p}\,.\label{twice}
\end{eqnarray}
Equations (\ref{relationA}) and (\ref{relationB}) for $M=\eta_{\xi \xi}$ are reported into
 (\ref{once}), which gives
\begin{equation}\label{etoile}
 \eta_\xi=\frac{1}{F^3}\eta_{\xi\xi t}\,.
\end{equation}
Then using (\ref{relationC}) for $M=\eta_{\xi \xi}$, and (\ref{twice}) and 
(\ref{etoile}), brings to the important relation
\begin{equation}\label{imprelation}
\eta_{\xi\xi p t} F^3=F^5\eta_{\xi\xi} -3\eta_{\xi\xi t}\eta_{\xi\xi p}.
\end{equation}
 The change of function (\ref{vary2},\ref{funcr2}) and the expressions of $F$, $F_p$ and $F_t$ give
\begin{equation} 
 r_{p t}=-\frac{1}{3}\left\{\frac{F^3 \eta_{\xi\xi p t} + 
 3\eta_{\xi\xi t}\eta_{\xi\xi p}}{F^6}\right\}.
\end{equation}
Finally using (\ref{imprelation}) and the expression of $F$ we obtain the Bullough-Dodd equation
\begin{equation} 
 r_{p t}=\frac{1}{9}\left\{\exp{\left(\frac{2r}{3}\right)}-\exp{\left(-\frac{r}{3}\right)}\right\}.
\end{equation}  
\label{sec:appendix}

\end{document}